\documentclass[aps,twocolumn,pre]{revtex4-1}
\usepackage{amsmath}

\usepackage{color}
\usepackage{amsfonts}
\usepackage{epsf}
\usepackage{graphicx}
\usepackage{natbib}
\usepackage{ulem}
\newcommand{\sign}{\text{sign}}
\definecolor{davide}{RGB}{3,101,3}
\definecolor{martin}{RGB}{51,153,255}
\definecolor{francesco}{RGB}{255,0,255}

\begin{document}

\title{Statistical mechanics of a single active slider on a fluctuating interface}

\author{F. Cagnetta, M. R. Evans,  D. Marenduzzo}

\affiliation{SUPA, School of Physics and Astronomy, University of Edinburgh, Peter Guthrie Tait Road, Edinburgh EH9 3FD, United Kingdom}

\begin{abstract}
We study the statistical mechanics of a single {\it{active slider}} on a fluctuating interface, by means of numerical simulations and theoretical arguments.
The slider, which moves by definition towards the interface minima, is active as it also stimulates growth of the interface.
Even though such a particle has no counterpart in thermodynamic systems, active sliders may provide a simple model for ATP-dependent membrane proteins that activate cytoskeletal growth. 
We find a wide range of dynamical regimes according to the ratio between the timescales associated with the slider motion and the interface relaxation.
If the interface dynamics is slow, the slider behaves like a random walker in a random envinronment which, furthermore, is able to escape environmental troughs by making them grow.
This results in different dynamic exponens to the interface and the particle: the former behaves as an Edward-Wilkinson surface with dynamic exponent $2$ whereas the latter has dynamic exponent $3/2$.
When the interface is fast, we get sustained ballistic motion with the particle surfing a membrane wave created by itself.
However, if the interface relaxes immediately (i.e., it is infinitely fast), particle motion becomes symmetric and goes back to diffusive.
Due to such a rich phenomenology, we propose the active slider as a toy model of fundamental interest in the field of active membranes and, generally, whenever the system constituent can alter the environment by spending energy.
\end{abstract}

\maketitle

\section{Introduction}

Active matter systems are collections of active particles, which consume energy, for example ATP, in order to perform work or modify their surroundings~\cite{ramaswamy2010rev}.
Their study lies at the interface between statistical, soft matter and biological physics.
An interesting example of active matter is realised when the active particles are embedded in a fluctuating membrane (as opposed to the usual scenario of active particles in a thermal fluid).
The resulting nonequilibrium system is termed an active membrane~\cite{ramaswamy2000actmem,gov2006dynamics,ben2011effective,maitra2014a,das2016aa} and here we investigate some of its general features.
We shall generically define an active membrane as a composite system consisting of a membrane with added active elements which we shall refer to as inclusions.
The archetypal instance of such a system is the plasma membrane of eukaryotic cells, where the membrane is the lipid bilayer, whereas the active inclusions are the so-called membrane proteins, which consume ATP to perform tasks such as proton pumping, ion channeling, or cytoskeletal polymerisation~\cite{fluidmosaicmembrane1,fluidmosaicmembrane2}. 

In~\cite{cagnetta2018aa} we introduced in a simple and generic model of an active membrane.
The motivation was two-fold: (i) to identify the key principles underlying pattern organisation at the leading edge of a moving cell and (ii) to test the applicability of Kardar-Parisi-Zhang (KPZ) scaling in an active setting.
In our model, active inclusions mimic Rho-like proteins~\cite{hall1998rho}, which stimulate the growth of interfacial, lamellipodium-like protrusion, while being, in turn, coupled the the membrane curvature.
Loosely speaking, their shape causes them to accumulate into membrane ``valleys'' (i.e. regions with negative curvature; see the Model section for more details on the mapping with biology). 
We found that the coupling between active elements and interface dynamics leads to an intriguing form of microphase separation, and to patterns, such as waves,  which cannot be realised within an equilibrium setting.
We also found that, due to energy input from active elements at the local level, the interface dynamics is not described by the KPZ universality class, but displays a novel oscillatory behaviour superimposed to an Edwards-Wilkinson dynamic scaling~\cite{cagnetta2018aa}. 

In this paper we take a step back and study the single-particle limit of the model of \cite{cagnetta2018aa}.
This allows us to focus on the effects of the inclusion-interface coupling, and eliminate the collective behaviour emerging from the interactions amongst many inclusions.
As we concentrate on the fundamental statistical mechanics of the system, we generically refer to our active particle as a ``slider'' (as it slides down interfacial height gradients).
The slider is active as it also promotes interfacial growth in contrast to the previously studied passive slider problem (see section \ref{ssec:passive}).
We will place particular emphasis on the role of the ratio between the characteristic timescales of slider and interface dynamics.

\begin{figure}[h!]
\begin{center}
  \begin{tabular}{cc}
       \includegraphics[width=1\columnwidth]{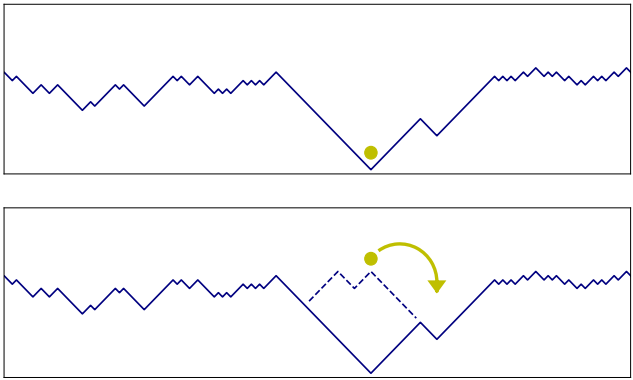}\\
  \end{tabular}
\caption{Pictorial representation of the system dynamics for $\omega\leq 1$, where $\omega$ is the interface-to-slider timescale ratio mentioned in the text. In the top panel, the particle has reached a local interface minimum. After some time (bottom panel), this location has become a local maximum due to local growth stimulation by the particle. The particle can now diffuse to the next local minimum, as suggested by the yellow arrow.}
\label{fig:OmLt1Cartoon}
\end{center}
\end{figure}

\begin{figure}[h!]
\begin{center}
  \begin{tabular}{cc}
       \includegraphics[width=1\columnwidth]{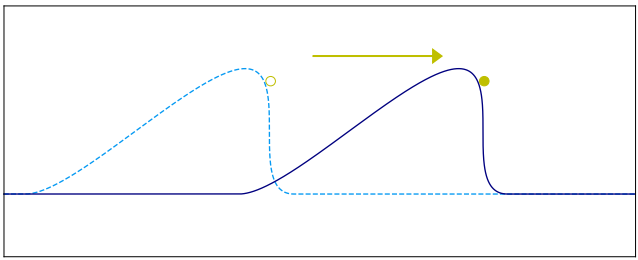}\\
  \end{tabular}
\caption{Cartoon depicting the system behaviour for intermediate $\omega$. The particle is surfing a membrane wave that is both pushing and being pulled by the particle.}
\label{fig:OmLargeCartoon}
\end{center}
\end{figure}

\begin{figure}[h!]
\begin{center}
  \begin{tabular}{cc}
       \includegraphics[width=1\columnwidth]{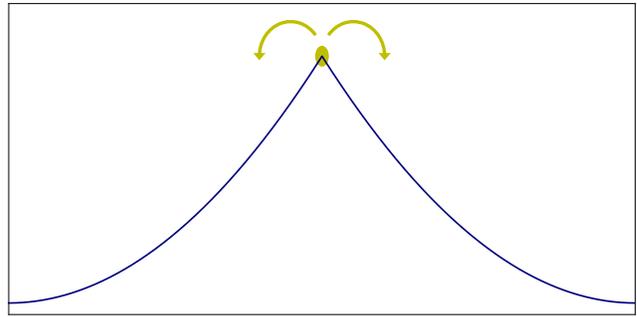}\\
  \end{tabular}
\caption{Typical interface shape for $\omega\rightarrow\infty$. As the interface attains a tent-like profile between two subsequent particle jumps, there is no longer a preferred direction for the particle to move along. Hence, it diffuses around as if it was not coupled to the interface.}
\label{fig:OmInfCartoon}
\end{center}
\end{figure}

In this work, our principal result is that the coupling between the active slider and membrane fluctuations yields three possible dynamical regimes, depending on the interface-to-slider timescale ratio.
First, if the slider diffusion is fast with respect to interface relaxation, the slider quickly reaches the closest valley, stimulates local interfacial growth until the valley becomes a peak, and then diffuses away to a neighbouring valley to repeat the process (Fig.~\ref{fig:OmLt1Cartoon}).
Viewing the interface profile as a free energy landscape, this process resembles {\it metadynamics}, a method of computational physics aimed at easing the sampling of complex free energy landscapes~\cite{laio2002aa}.
Second, if the slider dynamics is slower than interfacial relaxation, we instead observe an intriguing {\it surfing} dynamics, whereby the membrane bump created by the slider travels ballistically and pushes the slider itself forward (Fig.~\ref{fig:OmLargeCartoon}).
Finally, for infinitely fast interfacial dynamics, the surfing particle regime gives way to a third regime, where fast local growth forces maximum curvature at the location of the slider, which is now undergoing purely diffusive motion (Fig.~\ref{fig:OmInfCartoon}).

The paper is organised as follows.
In Section~\ref{sec:2} we describe the model, together with the observables we will measure; we also review the scaling theory within which we will formulate our results.
In Sections~\ref{sec:3}, \ref{sec:4} and \ref{sec:5}, we describe results in the three annouced regimes (with a small, large and intermediate interface-to-slider timescale ratio, respectively).
Notably, in \ref{sec:4} we show how the steady state dynamics of the corresponding regime reduces to an electrostatic problem where the active slider can be seen as a charged particle with the interface playing the role of the electrostatic potential.
In Section~\ref{sec:5}, we propose an analytic description of our system in terms of a set of Langevin equations for the moving slider position and interface height field, and we conclude in Section~\ref{sec:conc}.

Let us first, however, review the contrasting case of passive scalar dynamics.

\subsection{Passive slider problem}\label{ssec:passive}

From the purely statistical mechanics point of view, the problem of a particle moving on a fluctuating interface is related to that of passive scalar dynamics, where one or more passive tracers are coupled to a generally far-from-equilibrium medium.
Such a system is realised, for instance, when fluorescent dyes are used to highlight turbulent flow in a fluid~\cite{kraichnan1994a}, with passive particles sliding down a fluctuating potential landscape~\cite{das2000aa,das2001aa,drossel2002aa}, or with a so-called second class particle, whose dynamics is designed to locate shocks in driven diffusive systems like ASEP~\cite{derrida1999aa}.

The agents in these problems are {\it passive}, in the sense that they do not affect the dynamics of the medium they are moving in.
A possible back-coupling has instead been considered in the context of sedimenting colloidal crystals~\cite{lahiri1997aa} and in biophysical models for membrane proteins diffusion~\cite{reister2005aa,leitenberger2008aa,naji2009aa,dean2011aa}.
Such a problem has been attracting interest since the advent of single-particle tracking techniques, with the improved experimental characterisation of membrane protein dynamics calling for an accurate theoretical description.
In~\cite{reister2005aa,leitenberger2008aa}, for instance, the authors analyse corrections to the protein 2D diffusion coefficient due to the membrane fluctuations in the third dimension, while works such as~\cite{naji2009aa,dean2011aa} consider the effect of the two-fold coupling between protein shape and membrane curvature.
All these studies, however, are limited to an equilibrium or quasi-equilibrium setting, whereby the coupling between particle and interface corresponds to the minimisation of some prescribed effective free energy.
In the case of protein-curvature coupling, for instance, proteins tend to sit on membrane regions with a given curvature, and they also tend to impose such a given curvature on the membrane region where they sit.

It is worth remarking that an extension towards genuinely out-of-equilibrium conditions of the model introduced in~\cite{lahiri1997aa} has been proposed and analysed in the series of papers~\cite{chakraborty2016a,chakraborty2017a,chakraborty2017b}.
The authors therein consider a mixture of heavier and lighter particles that can stimulate the interface in different ways, e.g. the heavier push it down whereas the lighter lift it up.
We, instead, focus on particles that pull the interface up but then slide down the resulting interfacial slope.

\section{The Model}\label{sec:2}

Our model describes a random walker coupled to a fluctuating landscape.
The latter is a 1+1-dimensional interface, whose configuration is specified by a set of stochastic variables $\left\{h_i\right\}$.
Each of these variables represents the height of the interface over the $i$-th site of a one-dimensional ring-like lattice of length $L$.
As in standard surface-growth models, the height variables obey the solid-on-solid condition $|h_{i+1}-h_i|=1$, which causes the landscape to look like the trajectory of a random walker~\cite{meakin1986aa}.
In addition, the interface fluctuates according to local  dynamics, i.e. troughs of the interface ($\vee$) transform into peaks ($\wedge$) and vice versa.
The transition $\vee\rightarrow\wedge$, which causes the height to increase, has corresponding rate $p_i^+$; the transition $\wedge\rightarrow\vee$, which causes the height to decrease, has corresponding rate $p_i^-$. 

While the landscape evolves according to the aforementioned dynamical rules, the particle simply jumps between the lattice sites, with rates depending on the state of the surrounding environment.
We call such rates $q^R_k$ (for a right jump) and $q^L_k$ (for a left jump), where $k$ is the current lattice coordinate of the particle.
The dependence of the jump rates on the particle position is due the particle-landscape coupling, and is designed so that the interfacial slopes bias the local jump rates towards the site with lower height (troughs).
On the other hand, the particle renders the growth event $\vee\rightarrow\wedge$ more likely than the reversed one $\wedge\rightarrow\vee$ on the site where it sits.
Specifically,%put \\[1ex]
\begin{equation}\label{eq:IntRates}
 p_i^{\pm} =  p\left(1\right.\left. \pm \lambda\delta_{i,k} \right),
\end{equation}
\begin{equation}\label{eq:PartRates}
 q^R_k = q\left(1 - \gamma\nabla h_k\right),  q^L_k = q\left(1 + \gamma\nabla h_k\right),
\end{equation}
where $\nabla h_k = (h_{k+1}-h_{k-1})/2$ is the height gradient seen by the particle and $\delta_{i,k}$ the Kronecker delta. 

According to Eq. (\ref{eq:IntRates}), the particle is perceived by the interface as a defect at site $k$ whereby the up/down symmetry of fluctuations is broken, and the direction of the symmetry breaking depends on the sign of $\lambda$.
In the kinetic interfaces theory language~\cite{barabasi1995fractal}, the interface dynamics is Edwards-Wilkinson-like (EW)~\cite{edwards1982aa} everywhere except at the particle site, where it is Kardar-Parisi-Zhang-like (KPZ)~\cite{kardar1986aa}.
The interface, in turn, affects the particle motion as a potential---the effective potential energy is $\gamma h$.

It is useful to briefly pause at this point and comment in some more detail on the biophysical relevance of the $\gamma$ term to the case of ATP-dependent membrane activators~\cite{cagnetta2018aa}.
Such a coupling is allowed by symmetry for a moving interface~\cite{cai1995}, where it arises naturally as a kinematic ``advective'' contribution.
In this biophysical context, a positive $\gamma$ as considered here and in~\cite{cagnetta2018aa} appears automatically when a collection of membrane activators lead to cytoskeletal growth, hence cellular motility~\cite{ramaswamy2000actmem,maitra2014a}.
It may also model effectively chemically-induced biases towards or against substances in the cell cortex or interior, as the $h\to -h$ symmetry is broken even for a stationary membrane (as the membrane would separate the cytoplasm from the cellular exterior).
Within our geometry (akin to the so-called Monge gauge~\cite{cai1995,ramaswamy2000actmem}, this coupling has also the same broad consequences of curvotaxis~\cite{gov2006dynamics,shlomovitz2007membrane,peleg2011aa}, as proteins accumulate in valleys or peaks.
Yet its form is fundamentally distinct, as curvotaxis means sensitivity to gradients in the {\it curvature}, rather than height as done here. 

Coming back to our model, it is clear that $\lambda$ and $\gamma$ measure the extent of the coupling in our system.
Having both of them greater than zero induces a negative feedback of the kind discussed in~\cite{cagnetta2018aa}, whereby the particle shapes the landscape in a way that then repels it, continuously creating structures it is then pushed away from.
Throughout the paper, we shall set $\lambda=\gamma=1$, so that the uphill rate in Eq.~(\ref{eq:PartRates}) becomes zero together with $p^-_k$, the rate of $\vee\rightarrow\wedge$ at the particle site.
This is to avoid extended crossovers from the passive limits $\lambda=0$ and $\gamma=0$ and focus on the main goal of this paper, which is to explore the effect of varying $\omega=p/q$ on the system behaviour.
\begin{figure}[h!]
\begin{center}
  \begin{tabular}{cc}
       \includegraphics[angle=-90,width=1\columnwidth]{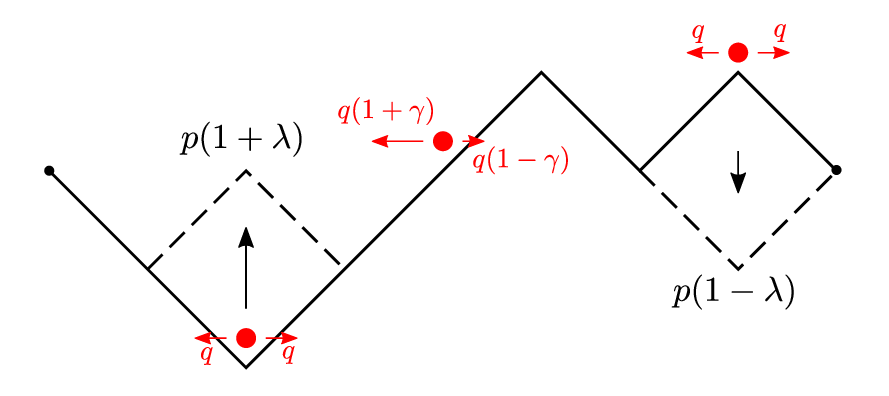}\\
  \end{tabular}
\caption{Schematics of our model active interface. The particle (red circle) presence favours the move that makes the interface grow, as denoted by the black arrows. The particle, in turn, will jum with left/right symmetric rates when on a hill or in a trough, as the leftmost and rightmost particles in the figure; whereas, when sitting on a slope as the middle particle of the figure, it will be more likely to jump towards the lower height.}
\label{fig:modelCartoon}
\end{center}
\end{figure}

The ratio $\omega$ measures how fast the interface dynamics is with respect to that of the particle.
Large $\omega$, for instance, implies that the interface dynamics is faster than that of the particle, and the interface adapts to the particle position before the latter moves significantly.
The converse is true for small $\omega$.
In order to efficiently change $\omega$ in simulations (see Supplementary Information of~\cite{cagnetta2018aa} for additional details), we use the following strategy.
Each timestep of the Monte Carlo (MC) algorithm consists of $N_s = aL + b$ micro-steps.
In each of the micro-steps, we choose the particle with probability $b/N_s$ or an interface site with probability $a/N_s$.
Once chosen, the particle has probability $\left(1 - \nabla h_X\right)/2$ of moving right and $\left(1 + \nabla h_X\right)/2$ of moving left.
For the interface updates, instead, first we check whether the chosen site is a peak or a trough. 
If the $i$-th site is a trough (peak) a local growth move is performed with probability $\left(1 + \lambda\delta_{X,i} \right)/2$ ($\left(1 - \lambda\delta_{X,i} \right)/2$).
The desired value of $\omega$ is thus selected by tuning $a$ and $b$, as their ratio measures the average number of updates of the interface per particle update.

In the remainder of this section we will define the observables of interest for the system at hand, and discuss their expected behaviour in relation to previous studies of similar problems.

\subsection{Observables and scaling}\label{ssec:2a}

As our system is made of two components (particle and interface), each pushing the other far from equilibrium, it is natural to characterise the dynamical and statistical properties of each component. 

In the theory of kinetic roughening, most of the global statistical properties of an interface can be discerned from its first two moments~\cite{plischke1984aa}, the mean height $\overline{h} = L^{-1}\sum_{i=1}^L h_i$ and the squared width $W^2 = L^{-1}\sum_{i=1}^L\left( h_i-\overline{h}\right)^2$.
Notice that both $\overline{h}$ and $W^2$ are stochastic variables, as are the $h_i$'s. 

We will denote the ensemble-averaged width (averaged over many realisations of the system dynamics) by lower case $w$.
The ensemble-averaged width is expected to follow the Family-Vicsek scaling hypothesis~\cite{family1985aa},
\begin{equation}\label{eq:WidthScalingHyp}
   w(L,t)  = L^{\alpha}f(t/L^{z_1}),
\end{equation}
where $\alpha$ and $z_1$ are the roughness and dynamic exponent of the interface, respectively, whereas the scaling function $f$ behaves as a power law for small arguments and a constant for large ones.
The width grows in time as a power law $\sim t^\beta$ until, at a time $t \sim L^{z_1}$, it saturates due to the finite interface  size.
According to Eq.~(\ref{eq:WidthScalingHyp}), the saturation value scales with the size as $L^{\alpha}$, while, in order to cancel any system size dependence at short times $t \ll L^{z_1}$, we must have $f(y) \sim y^{\alpha/z_1}$ for small $y$, which implies that the initial growth exponent obeys $\beta =\alpha/z_1$.

Following this line of thought, we will analyse the particle dynamics by looking at the first two moments of the particle displacement $X_t$.
As there is only one particle, averages here are performed over realisations of the stochastic dynamics.
Contrary to the height first moment, the average displacement of the particle is identically zero, as nothing breaks the left-right symmetry of averages (we will see though that such symmetry is broken at the individual trajectory level).
The mean squared displacement, however, obeys a scaling form akin to that of Eq.~(\ref{eq:WidthScalingHyp}),
\begin{equation}\label{eq:DiffScalingHyp}
  \left\langle X_t^2 \right\rangle = tL^{\chi}g(t/L^{z_2}),
\end{equation}
where $z_2$ is a dynamic exponent relating the time it takes for the particle to reach its steady-state behaviour to the system size.
The form of (\ref{eq:DiffScalingHyp}) can be understood from the requirement that at long times, on a finite system ($t\gg L^{z_2}$), the behaviour of the particle becomes diffusive $\left\langle X_t^2 \right\rangle \sim t$.
Thus the scaling function $g$ must be constant for large arguments, and $\chi$ specifies the system-size-dependence of the effective, long-time diffusion coefficient.
On the other hand, the early-time behaviour should not depend on the system size, and the small argument behaviour of $g$ is  fixed by requiring a functional form $g(y)\sim y^{\chi/z_2}$ which causes the $L$'s in Eq. (\ref{eq:DiffScalingHyp}) to cancel each other for $t\ll L^{z_2}$. 
Then one obtains the early times law 
\begin{equation}
\left\langle X_t^2 \right\rangle \sim  t^\eta,
\label{et}
\end{equation}
where 
\begin{equation}
\eta = 1 + \frac{\chi}{z_2}.
\label{etadef}
\end{equation}
The scaling hypothesis  Eq.~(\ref{eq:DiffScalingHyp}) was propsed in~\cite{derrida1999aa} for a `second class particle' which exhibits superdiffusive behaviour and was later used in related problems of Brownian particles passively coupled to time-dependent random environments \cite{chin2002aa,gopalakrishnan2005aa,nagar2006aa}.

Now, the theory of transport in random environment~\cite{bouchaud1990aa} states that the spatial correlations of a stochastic medium  may give rise to anomalous diffusion of the particles living there. Then one may write (for a
system of infinte spatial extent)
\begin{equation}
\left\langle X_t^2 \right\rangle \sim t^{2/z_P}
\end{equation}
where $z_p$ is yet another dynamical exponent.
It  characterises the anomalous diffusion as follows: after time $t$ the particle will have explored a distance $\left\langle X_t^2 \right\rangle^{1/2}  \sim t^{1/z_P}$.
Thus the particle should explore a finite  system size $L$ after time $t\sim L^{z_p}$.

However, {\it a priori}, the value of $z_p$ is not necessarily equal to that of $z_2$. 
Demanding that the two dynamical exponents $z_2$ and $z_p$ are indeed equal implies the scaling relation
\begin{equation}\label{eq:HyperscalingRel}
\chi + z_2 = 2\;.
\end{equation}
Such a special condition can be perceived as the signature that no other lengthscale than the system size affects the particle motion~\cite{chin2002aa}.
In fact, $\chi + z_2=2$ holds in the several ``passive'' versions of our model considered in the literature, such as the second class particle problem and that of a passive slider on a self-affine interface~\cite{derrida1999aa,bohr1993aa,chin2002aa}.
It appears, in addition, that also $z_1$ and $z_2$ can be identified with each other, at least in most of the problems we refer to~\nocite{singha2018aa}\footnote{The only exception seems to be the ``KPZ anti-advection'' case, as hinted in~\cite{drossel2002aa} and suggested by the numerics in~\cite{singha2018aa}.}.
Consequently, one single dynamic exponent suffices to characterise all dynamical features of the system.
We will soon see that it is not always the case in our active model.

\section{Fluctuating metadynamics at $\omega\leq 1$}\label{sec:3}

\begin{figure}[h!]
\begin{center}
  \begin{tabular}{cc}
       \includegraphics[width=1\columnwidth]{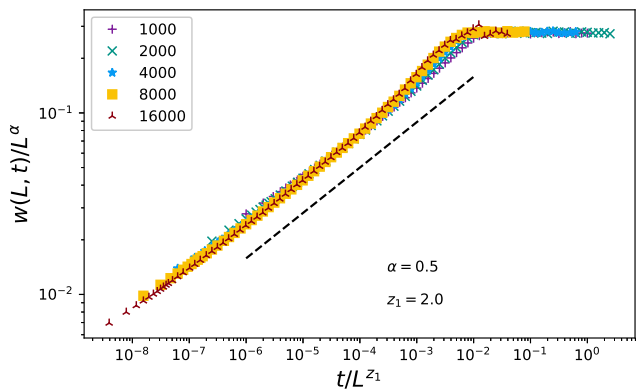}\\
  \end{tabular}
\caption{Scaling of the averaged width for $\omega=1$ -- with values of $L$ given in the legend. The best collapse is achieved by setting the exponents $\alpha$ and $z_1$ to the EW class values, even if the width oscillations cause a slight departure from the scaling hypothesis, Eq.~(\ref{eq:WidthScalingHyp}). A power law $x^{\alpha/z_1}$ is shown as a guide to the eye (black dashed line). The interface width has been averaged over at least $1000$ realisations of the stochastic dynamics for $L$ up to $8000$, and over $100$ realisations for $L=16000$. The number of realisations used for averages is the same in all the following figures, unless stated otherwise.}
\label{fig:Om1WidthScaling}
\end{center}
\end{figure}

Let us begin by setting $\omega=1$, i.e. considering the case where particle and interface have the same mobility.
On the $\gamma=\lambda$ line of the phase diagram, we find that the steady-state interface is described by Edwards-Wilkinson statistics.
Simulations (see Fig. \ref{fig:Om1WidthScaling}) in which the width is measured as a function of time show that the roughness exponent $\alpha = 1/2$ and the dynamic exponent $z_1=2$---the values of the EW class.
Note that the numerical estimation of the exponents is hampered by the emergence of width oscillations.
These oscillations begin to emerge for the larger system sizes in Fig. \ref{fig:Om1WidthScaling} just before saturation,  although the effect is not as strong as for the finite particle density case~\cite{cagnetta2018aa}, where oscillations are clearer and extend over several periods.

\begin{figure}[h!]
\begin{center}
  \begin{tabular}{cc}
       \includegraphics[width=1\columnwidth]{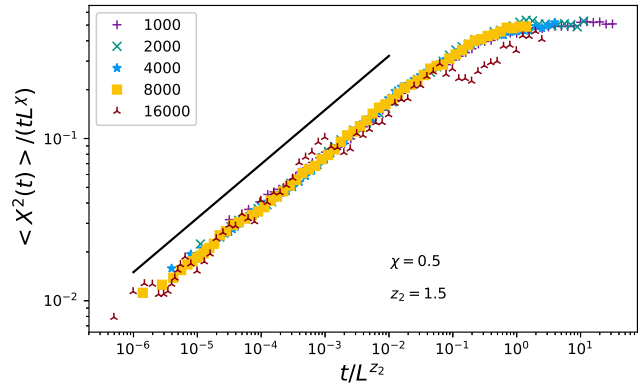}\\
  \end{tabular}
\caption{Scaling of the particle MSD for $\omega=1$, $L$ as in the key. The data are collapsed using values $\chi=1/2$ and $z_2=3/2$  consistent with the second class particle scaling discussed in the text. The black dashed line is a guide to the eye  suggesting $\left\langle X_t^2 \right\rangle \sim t^{4/3}$. In this figure, and all the following figures, the MSD is computed in steady state, meaning that time starts running after the interface has reached its saturation width.}
\label{fig:Om1MSDScaling}
\end{center}
\end{figure}

With regard to the particle MSD, the numerics agree with the scaling form in Eq.~(\ref{eq:DiffScalingHyp}), as is shown in Figure~\ref{fig:Om1MSDScaling}.
The exponents, $\chi = 1/2$ and $z_2 = 3/2$, are the same as in the second class particle problem \cite{derrida1999aa}, which, due to the well-known mapping between the totally antisymmetric simple exclusion process and a discrete interface model in the KPZ class, corresponds to setting $p^{\pm} = p(1\pm \lambda)$ uniformly over the interface instead of on the particle site only.
The exponent $z_2=3/2$, there, reflects the dynamic exponent of the interface and  the value $\chi = 1/2$ yields $\eta$ through (\ref{etadef}).
In this case the exponents obey the scaling relation (\ref{eq:HyperscalingRel}). 

In our model, conversely, there is a mismatch between $z_1$ and $z_2$, i.e. the interface and particle dynamic exponents are not the same.
A possible explanation for such a difference is the following. The exponent $z_1$ refers to the saturation of a global interfacial variable such as the width: it is reasonable to expect a single particle not to dramatically alter its properties.
The interface dynamics is thus dominated by the up/down-symmetric growth events away from the particle, resulting in $z_1=2$.
The value of $z_2$, on the other hand, is related to the early-time superdiffusive behaviour of the single particle (\ref{et}).
Such behaviour is triggered by the local environment of the particle rather than the instantaneous global structure.
Here, owing to the particle itself, the up/down symmetry of fluctuations is broken and the dynamic exponent $3/2$ is plausible.

In order to corroborate the idea that the particle experiences  a different dynamic exponent to that of  the interface as a whole, we measured the spatial spreading of correlations from the particle site. Specifically, we put a static ($q^L = q^R =0$), yet active particle (which still catalyses the interface growth) on the $k$-th site of a flat interface, then let the interface evolve and measure the slope correlation function
\begin{equation*}
  C_s(j,t) = \left\langle \left(h_{k+1}(t)-h_k(t)\right) \left(h_{k+j+1}(t)-h_{k+j}(t)\right) \right\rangle
\end{equation*}
at different times.
The average here is performed over several histories of the interface dynamics and, due to the left-right symmetry, we limit our measurements to the half of the interface on the right of the particle.
\begin{figure}[h!]
\begin{center}
  \begin{tabular}{cc}
       \includegraphics[width=1\columnwidth]{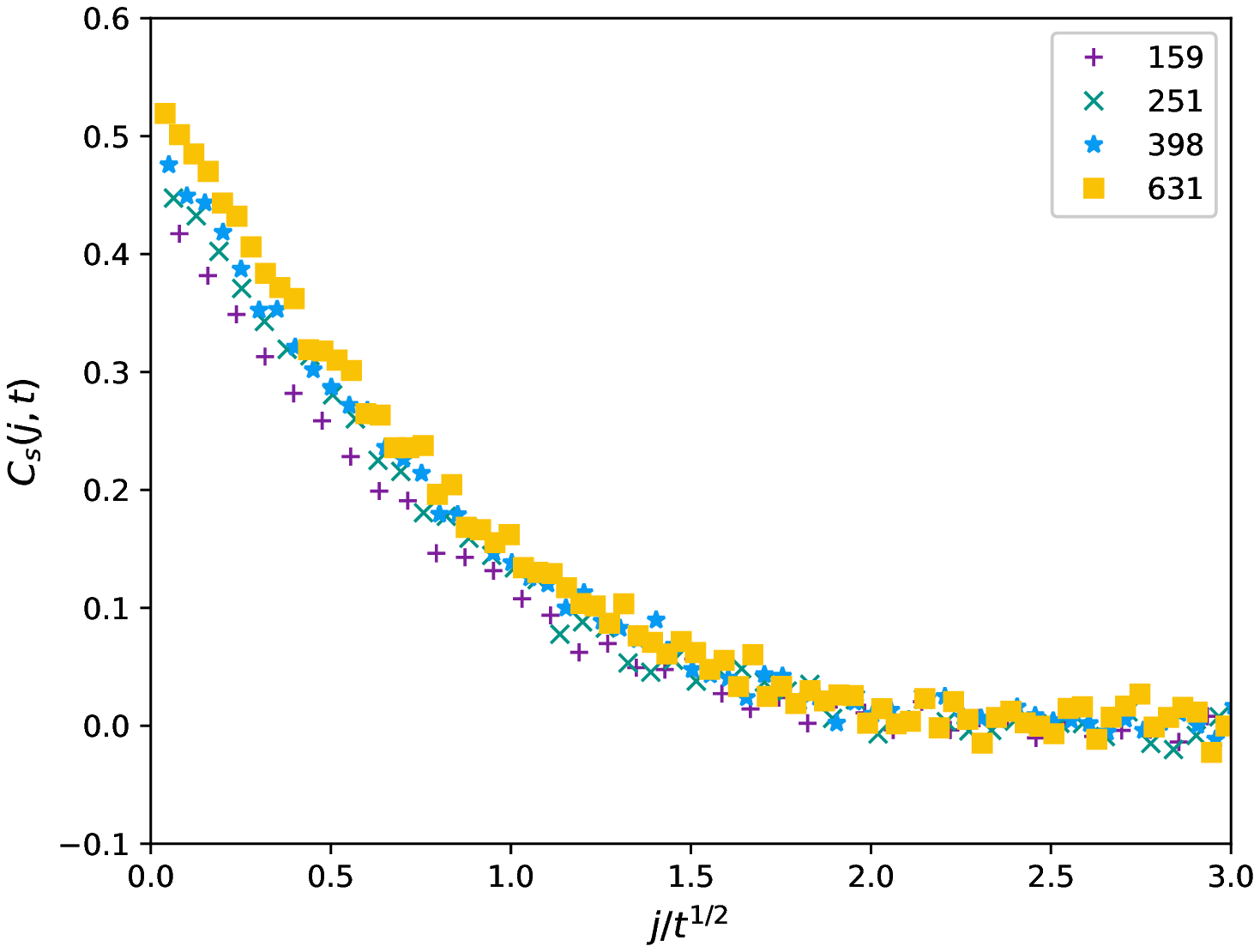}\\
       \includegraphics[width=1\columnwidth]{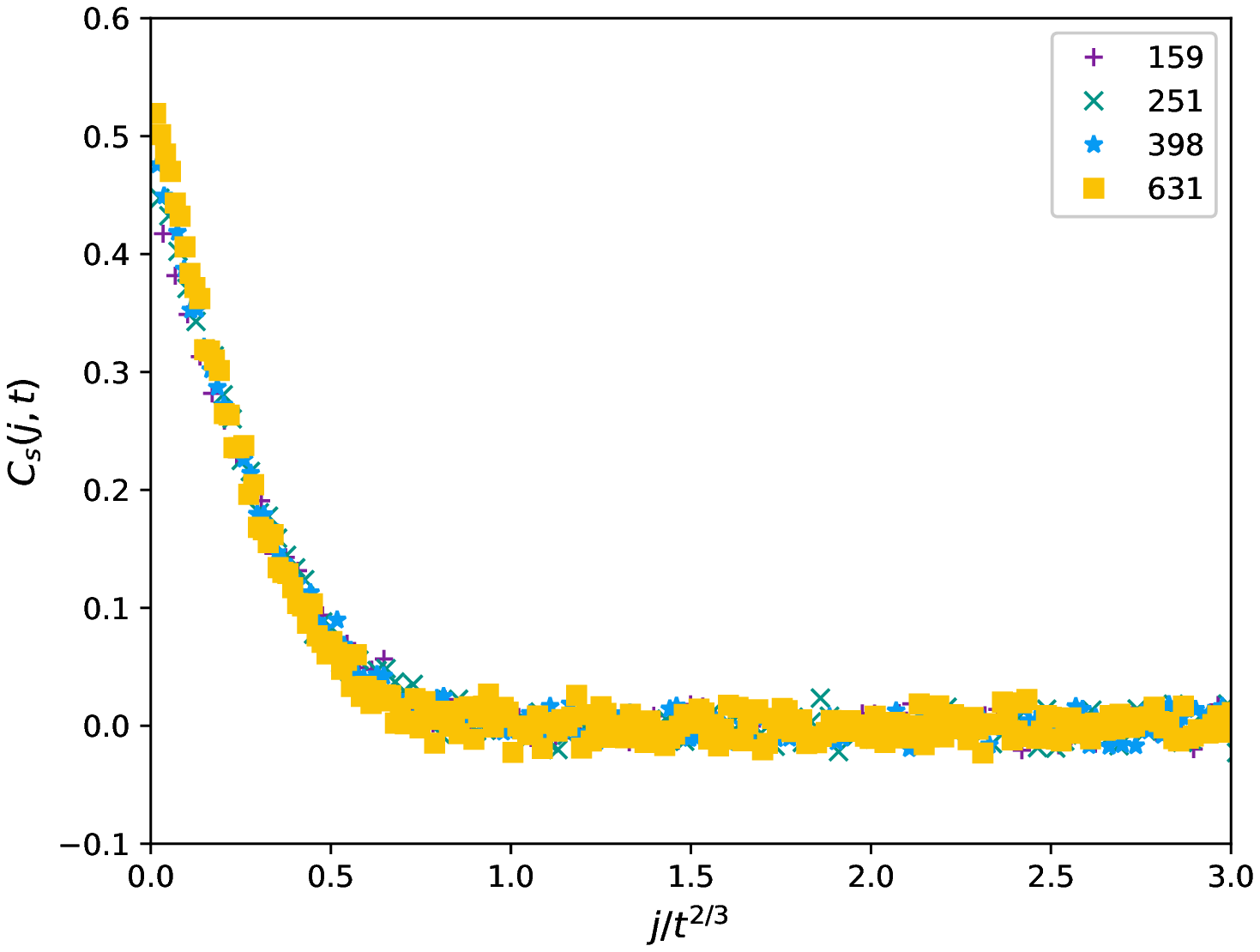}\\
  \end{tabular}
\caption{Correlations spreading from  a fixed active particle which catalyses growth in the interface. The averages here are performed over $10000$ different realisations of the interface dynamics. The slope correlation function defined in the text is plotted against $j/t^{1/2}$ on the top panel and $j/t^{2/3}$ on the bottom panel. The overlap of the functions is much better in the latter case, suggesting that correlations spread around the particle as $t^{1/z_2}$ whwre $z_2= 3/2$.}
\label{fig:CorrSpreading}
\end{center}
\end{figure}
The data collapse of Fig.~\ref{fig:CorrSpreading} provides evidence that around the particle the correlation length grows, at least for relatively short times, as $t^{1/z_2}$  with $z_2=3/2$.
This is consistent with the dynamical exponent of the KPZ universality class.

To summarise the dynamics in the $\omega=1$ case, the interface behaves as an EW one with emerging  oscillations  analogous to those previously observed in the finite particle-density system.
The particle, in turn, behaves as if it were passively sliding on a KPZ interface, displaying an initial superdiffusive regime $\left\langle X_t^2 \right\rangle \sim t^{4/3}$, followed by normal diffusion $\left\langle X_t^2 \right\rangle \sim D_{\rm eff} t$ with $D_{\rm eff}\sim L^{1/2}$.
The crossover, caused by the system finite size, occurs at a time $t\sim L^{3/2}$.
Our independent measurement of correlations supports the idea that such a KPZ-like scaling is caused by the local, symmetry-breaking action of the particle, which causes itself to see the globally EW interface as an effectively KPZ one. 

\subsection{Activity versus trapping in the small $\omega$ limit}\label{ssec:3a}

\begin{figure}[h!]
\begin{center}
  \begin{tabular}{cc}
       \includegraphics[width=1\columnwidth]{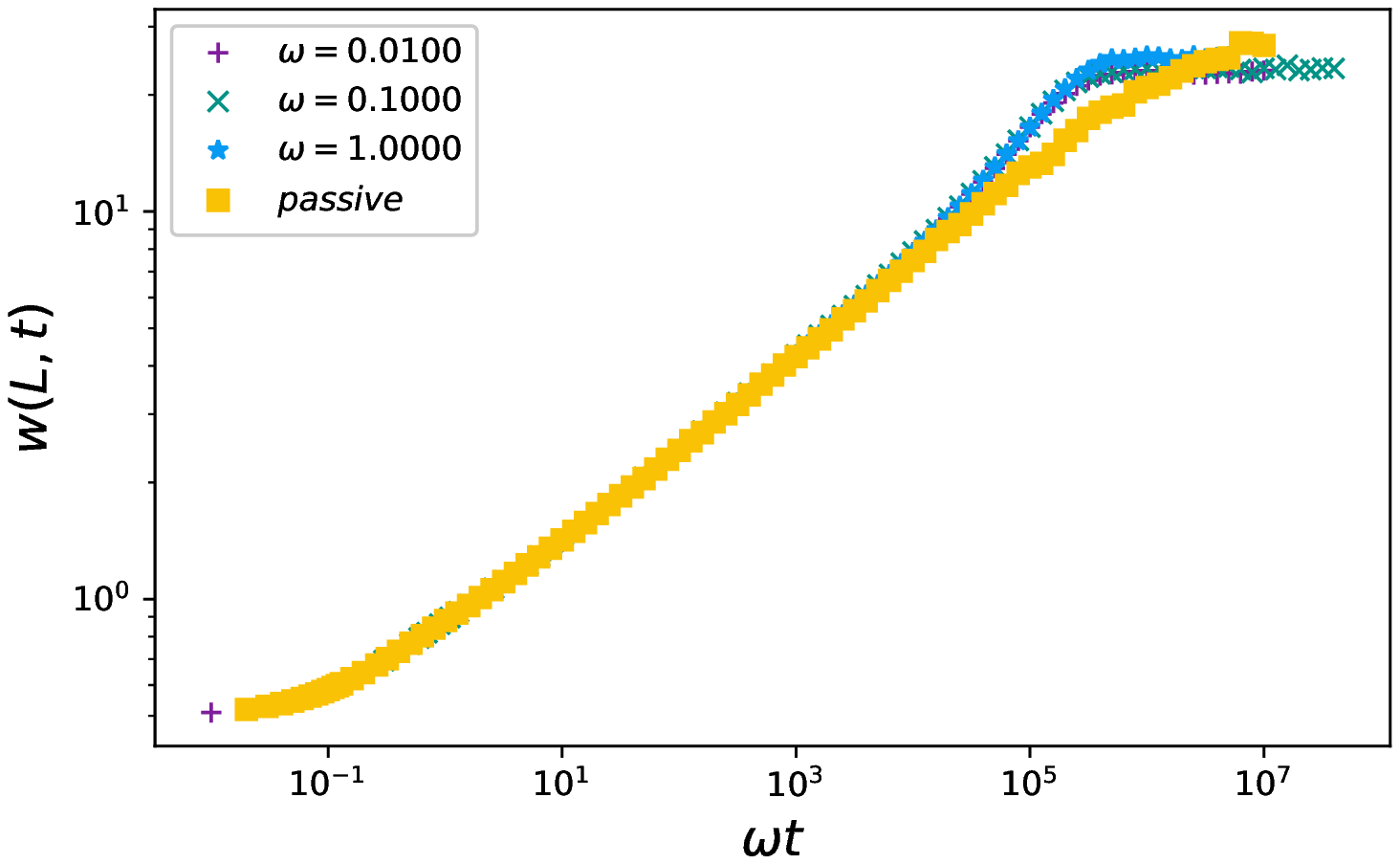}\\
       \includegraphics[width=1\columnwidth]{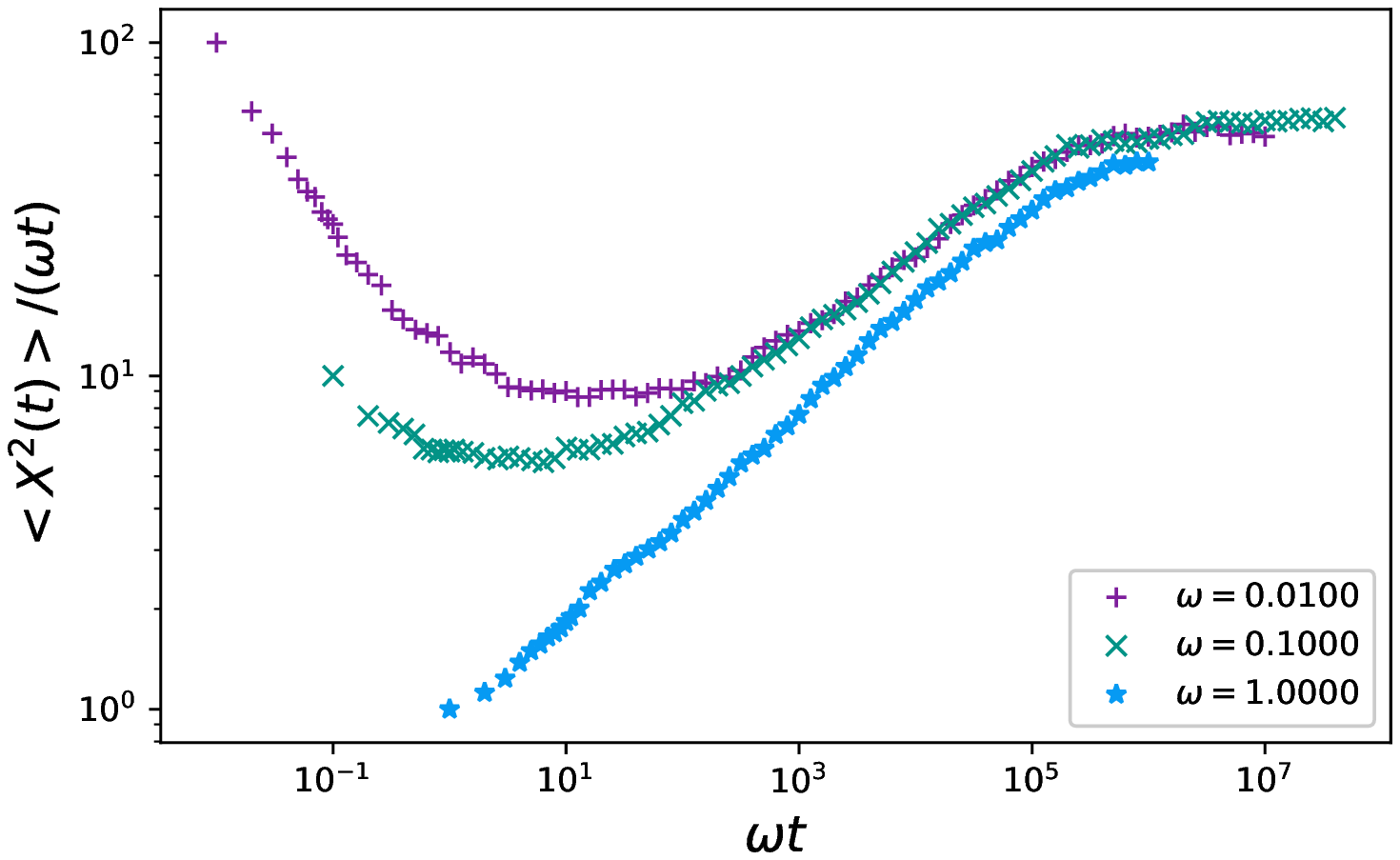}\\
  \end{tabular}
\caption{Width (left) and MSD (right) vs $\omega t$, for $L=8000$ and $\omega$ as in the key. The width of a passive, EW interface is also shown for comparison. While rescaling time by $\omega$ renders the interface dynamics independent of this parameter, the particle displays an early-time, subdiffusive regime, the extent of which scales as $1/\omega$, as pointed out in the text.}
\label{fig:ScalingOm}
\end{center}
\end{figure}
In this section we discuss the scaling of the interface width and of the particle MSD when the latter is slower than the former, i.e. $\omega <1$.
Let us  start by comparing, for $L$ fixed, the average width and MSD of systems with different $\omega$.
As shown in Figure~\ref{fig:ScalingOm}, upper panel, the width dynamics does not depend on $\omega$, apart from a trivial rescaling of time---recall that $\omega$ equals the average number of interface updates per particle update, whereas we take the average time for a particle move as our unit of time.
The interface exponents will then be the same as those observed at $\omega=1$, that is the EW class values $\alpha=1/2$ and $z=2$.

\begin{figure*}[ht!]
\begin{center}
  \begin{tabular}{cc}
       \includegraphics[width=1\columnwidth]{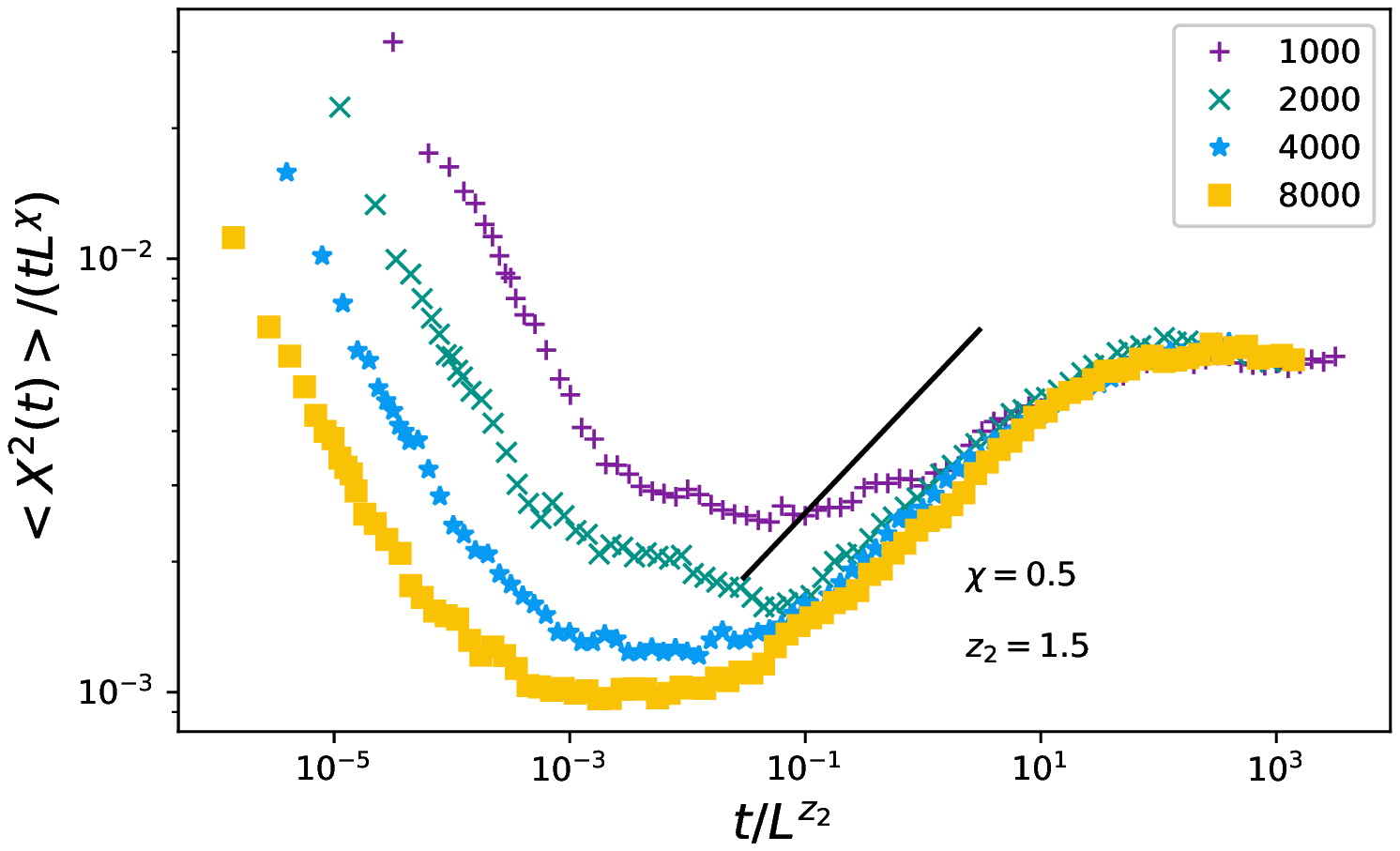} & \includegraphics[width=1\columnwidth]{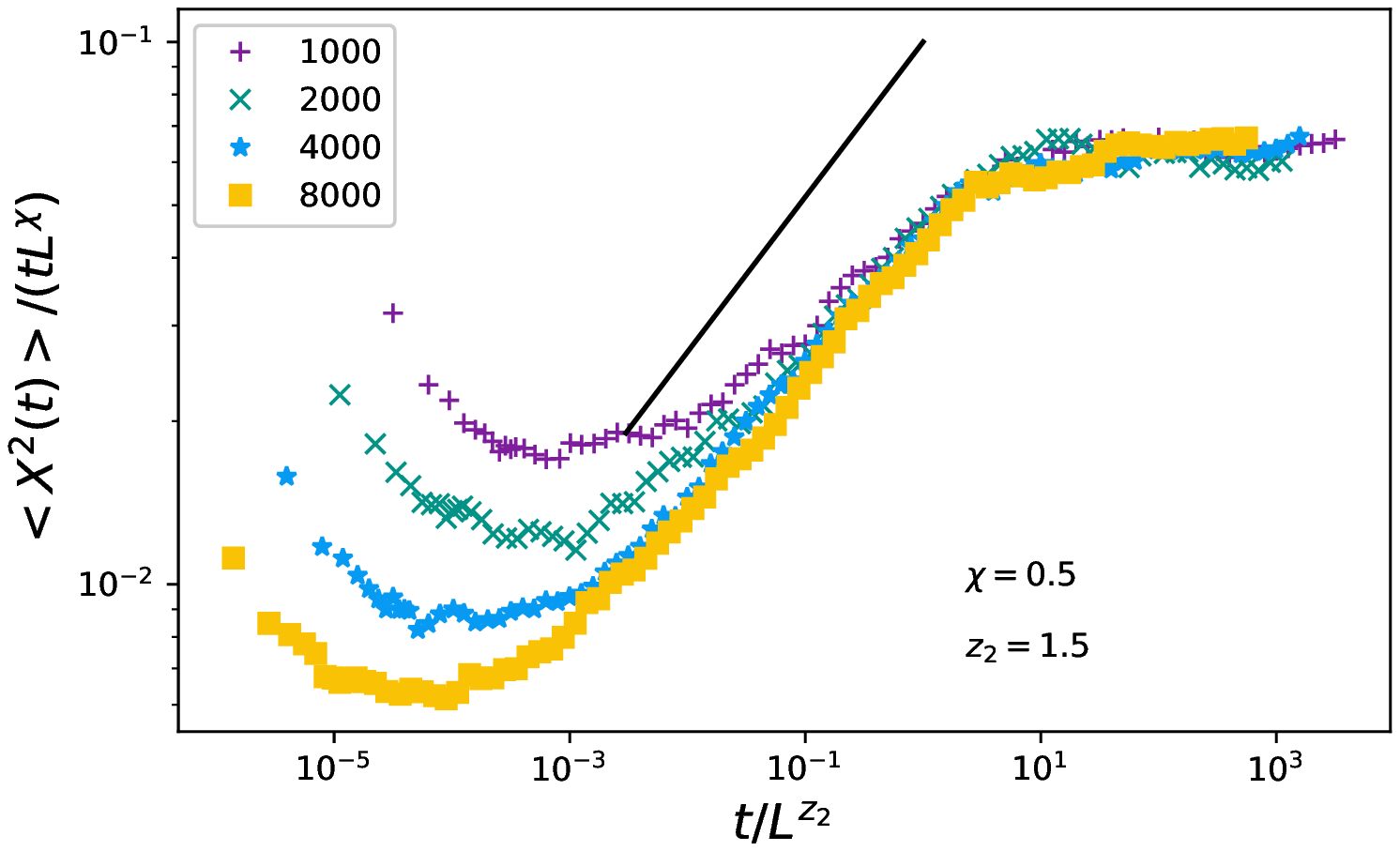}\\
  \end{tabular}
\caption{Scaling plot of the particle MSD for $\omega=10^{-2}$ (left) and $10^{-1}$ (right). The black solid lines in both plots are guides to the eye for the superdiffusive law $\left\langle X_t \right\rangle \sim t^{4/3}$.}
\label{fig:OmLT1MSDScaling}
\end{center}
\end{figure*}
The MSD, conversely, shows an initial subdiffusive regime, reminiscent of the typical behaviour displayed by random walkers in random environments. 
Subdiffusive behaviour is manifest in Figure~\ref{fig:ScalingOm}, lower panel, by a decreasing curve when $\left\langle X^2_t \right\rangle/(\omega t)$ is plotted as a function of $\omega t$.
In a completely static random environment (such as a quenched random potential) the sub diffusive behaviour due to trapping  can be as slow as $\left\langle X^2_t \right\rangle \sim (\log{t})^4$~\cite{sinai1983aa}. 
After a time  $\sim \omega^{-1}$, Figure~\ref{fig:ScalingOm}, lower panel, reveals that subdiffusion is replaced by superdiffusion, which  eventually crosses over to normal diffusion ($\left\langle X^2_t \right\rangle/(\omega t)  \to const.$ ) due to the finiteness of the medium, as in the $\omega=1$ case.
As a result, the scaled MSD at $\omega <1$ tends to the $\omega=1$ curve for sufficiently large scaled times, as in the right panel of Fig. \ref{fig:ScalingOm}.
In fact, $\omega^{-1}$ is the average time at which the interface site under the particle undergoes its first update. Hence, this is the time at which the interface activity steps in, together with the mechanisms responsible for the physics of the system at $\omega=1$.
The asymptotic properties of the system are then described by the same exponents found at $\omega=1$, it will just take longer times and larger systems for these to appear.
\begin{equation}\label{eq:OmLt1Exponents}
 \omega \leq 1: \qquad\begin{aligned}\alpha &= 1/2,\quad z_1=2;\\ \chi &= 1/2,\quad  z_2=3/2.\\ \end{aligned}
\end{equation}

We close this section by showing the MSD scaling at fixed $\omega$, with $L$ in the range $1000$---$8000$, so as to support the hypothesis $\chi=1/2$, $z_2=z_p=3/2$ in the whole $\omega\leq 1$ of the parameter space (Fig. \ref{fig:OmLT1MSDScaling}).
Even though there is no early time collapse of the  curves for different $L$, the late time collapse of the superdiffusive and diffusive regimes is fully compatible with the proposed exponents.
In order to obtain cleaner scaling behaviour, one would like  the sub- and super-diffusive regimes to be well separated in time.
Such a separation, however, would require system sizes much bigger than those used throughout this paper, hence significantly longer simulations---as we are dealing with a single particle on a fluctuating interface, for each particle trajectory one needs to simulate the whole interface dynamics too.

\section{The $\omega\rightarrow\infty$ limit: electrostatics on the ring}\label{sec:4}

Having explored the dynamics for $\omega\leq 1$, we now turn  to $\omega >1$, where the interface moves faster than the particle.
Let us start by considering the extreme case, i.e. the $\omega\rightarrow\infty$ limit.
The interface, in this limit, reaches a stationary state before the particle can even move: most of the system features, including the various exponents considered in this paper, depend on the stationary shape the interface reaches between subsequent jumps of the particle, as such a shape will determine the particle jump rates in the following step.

\begin{figure}[h!]
\begin{center}
  \begin{tabular}{cc}
       \includegraphics[width=1\columnwidth]{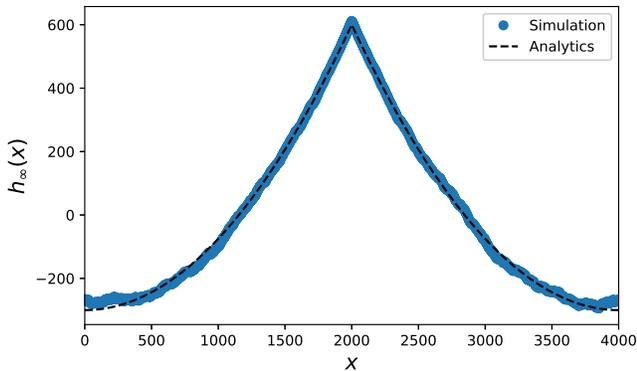}\\
  \end{tabular}
\caption{Tent-like shape of the interface, which is obtained in our model in the limit $\omega\rightarrow\infty$. The dotted line plots the analytic prediction (\ref{eq:OmInfIntDeterministic}), with $\Lambda/\nu=2$ and $L=4000$, while the blue dots represent a simulated $L=4000$ interface.}
\label{fig:OmInfInt}
\end{center}
\end{figure}

In order to gain insight into this stationary shape, we consider the related problem of a stochastic interface with a defect site.
Forcing our particle to stay put on a single site  means that the interface is being pulled from this specific site, whilst there are only up/down-symmetric fluctuations elsewhere on the ring.
By mapping our interface problem onto a simple exclusion process, one can readily infer what the system steady state is, especially with our parameter choice $\lambda=1$.
Imagine every $-1$-slope segment of the interface to be a bead and each $+1$-slope one to be a hole: then the transition $\vee \rightarrow \wedge$ (resp. $\wedge \rightarrow \vee$) corresponds to a bead moving right (resp. left).
All the beads in our system move left or right at the same rate and experience hard-core repulsion, whereas the particle site acts as a semi-permeable membrane which allows the beads to cross it only from the left to the right.
All the beads starting on the membrane left will eventually cross it and lie on its right, while there will be only holes on the left.
As a result, the steady-state interface will look like a ``tent" - a macroscopic convex wedge as depicted in Fig. \ref{fig:OmInfInt}.

We now consider fluctuations of the system.
We use a field-theoretic representation of the interface using symmetry considerations to retain the relevant terms.
The starting point is the Edwards-Wilkinson equation,
\begin{equation}\label{eq:EW}
 \partial_t h = \nu \nabla^2 h + \sqrt{2\Delta}\eta(x,t),
\end{equation}
where $h(x,t)$ is the interface field and $\eta$ a space-time white, Gaussian noise with unit variance.
In Eq.~(\ref{eq:EW}), $\nu$ can be perceived as the interface tension, while $\Delta$ is the noise intensity.
As we are writing down a field description from symmetry considerations, there is no explicit link at this level between the parameters in Eq.~(\ref{eq:EW}) and our stochastic model parameters.
This is however irrelevant for what we are going to deduce from the field-theoretic approach.
Periodic boundary conditions are used to enforce the ring topology on the system. 

The defect site (we will call its position $X_0$) is introduced as a $\delta$-like source term in the right-hand side of Eq.~(\ref{eq:EW}): 
\begin{equation}\label{eq:EW+defect}
 \partial_t h = \Lambda\delta(x-X_0) + \nu \nabla^2 h + \sqrt{2\Delta}\eta(x,t),
\end{equation}
where $\Lambda$ measures the strength of the bias on the defect site~\footnote{The microscopic rules in our model, actually, cause the interface fluctuations to occur normally to the interface local tangent: the $\delta$ source should therefore, in principle, be projected onto the local tangent. We will, instead, neglect such projection, as the resulting additional terms in the field equation would anyway be confined to the defect site, whereby the $\delta$ source makes any other addition practically irrelevant.}.
To infer the steady state of Eq. (\ref{eq:EW+defect}), we introduce the height Fourier modes $h_k(t) = \int_0^L dx\, h(x,t) e^{-ikx}$, where, due to periodicity, $k=2\pi n/L$ with $n$ integer.
Equation (\ref{eq:EW+defect}) transforms to
\begin{equation}
\label{eq:EWk}
 \partial_t \tilde h_k = \Lambda{\rm e}^{-ik X_0} - \nu k^2 \tilde h_k + \sqrt{2\Delta}\tilde \eta_k\;.
\end{equation}
Each mode is nothing but an Ornstein-Uhlenbeck process with $k$-dependent parameters.
The new ingredient here over the usual EW equation is the constant forcing term $\Lambda{\rm e}^{-ik X_0}$ which stems from the defect site.
Integrating (\ref{eq:EWk}) yields
\begin{equation}\label{eq:heightModes}
\begin{aligned}
 \tilde{h}_k(t) &= e^{-\nu k^2 t}\tilde{h}_k(0)\\
 &+ \int_0^t ds\,e^{-\nu k^2(t-s)}\left[ \sqrt{2\Delta}\tilde{\eta}_k(s) + \Lambda e^{-ikX_0}\right].
\end{aligned}
\end{equation}
First, by setting a flat initial condition, the $\tilde{h}_k(0)$ contribution disappears.
Next, we note that the $0$-th mode, once divided by $L$, equals the mean height $\overline{h}$ defined in section \ref{ssec:2a}; the mean height is a Gaussian random variable with mean $\Lambda t/L$ and variance $2\Delta t/L^2$.
The $0$-th mode contributes neither to the interface width, since this  measures fluctuations about the mean, nor to the particle dynamics, which only involves the interface slope.
We therefore ignore the zero mode for the remainder of this section.
In particular, we consider the stationary state
\begin{equation}\label{eq:OmInfInt}
 \lim_{t\rightarrow\infty}\left( h(x,t) - \overline{h} \right) = h^{\rm det}(x) + h^{\rm rand}(x), 
\end{equation}
where $h^{\rm det}(x)$ is the deterministic part of the interface profile coming from the delta function and Laplacian in (\ref{eq:EW+defect}) whereas $h^{\rm rand}(x)$ is a random part coming from the noise. 

We note in passing that the deterministic part of the steady state solution of (\ref{eq:EW+defect}) corresponds to the solution of Laplace's equation for the electrostatic potential of a point charge at $X_0$ on a one-dimensional lattice with periodic boundary conditions.
In order to have a consistent equation, which satisfies the periodic boundary condition, one has to introduce a background charge density to give overall charge neutrality.
In our context, this procedure  is equivalent to substracting out the $0^{\rm th}$ mode.

Let us then consider the $t\rightarrow\infty$ limit of all the other modes.
In this limit the deterministic part of (\ref{eq:heightModes}) reads
\begin{equation}\label{hkss}
\tilde h^{\rm det}_k(x) =  \lim_{t\rightarrow\infty} \frac{\Lambda}{\nu}\frac{1-e^{-\nu k^2t}}{k^2}e^{-ikX_0} = \frac{\Lambda}{\nu}\frac{e^{-ikX_0}}{k^2}.
\end{equation}
Inverting (\ref{hkss}) by  summing over $k=2\pi n/L$, $n\neq 0$, one gets
\begin{equation}
 h^{\rm det}(x) \equiv \frac{\Lambda}{\nu}\frac{L}{4\pi^2}\left[{\rm Li}_2(e^{2\pi i\frac{(x-X_0)}{L}})+{\rm Li}_2(e^{-2\pi i\frac{(x-X_0)}{L}})\right],
\end{equation}
where ${\rm Li}_m(x) = \sum_{k\geq 1} x^k/k^m$ is the polylogarithm of order $m$ of $x$.
We now invoke  a polylogarithm identity 
\begin{equation}
{\rm Li}_n(e^{2\pi i x}) + (-1)^n {\rm Li}_n(e^{-2\pi i x}) = -(2\pi i)^n B_n(x)/n!\quad  
\end{equation}
where  $0< x \leq 1$. Here, $B_n$ denotes the $n$-th Bernoulli polynomial and in particular, $B_2(x) = x^2-x+1/6$ so that we obtain
\begin{equation}\label{eq:OmInfIntDeterministic}
 h^{\rm det}(x) = \frac{\Lambda L}{2\nu} \left[\frac{(x-X_0)^2}{L^2}-\frac{|x-X_0|}{L}+\frac{1}{6}\right].
\end{equation}
This is the tent profile illustrated by the dashed line in Figure~\ref{fig:OmInfInt}.

The stochastic contribution $h^{\rm rand}(x,t)$  consists of  zero-average Gaussian random variables at each point $x$ of the profile and the Fourier transform is
\begin{equation}
 \tilde{h}^{\rm rand}_k(t) =   \int_0^t ds\,e^{-\nu k^2(t-s)} \sqrt{2\Delta}\tilde{\eta}_k(s).
\end{equation}
Inverting the Fourier transform by a  calculation analogous to that presented above reveals that, as $t\to \infty$,
\begin{equation}
h^{\rm rand}(x) \to \frac{\sqrt{2\Delta}}{L} \sum_{k\neq 0} \frac{{\rm e}^{ikx}}{\nu k^2} \tilde{\eta}_k,
\end{equation}
where 
\begin{eqnarray}
\langle \tilde{\eta}_k \rangle &=& 0\\
\langle \tilde{\eta}_k \tilde{\eta}_{k'}\rangle &=& L \delta(k+k')
\end{eqnarray}
so that $\langle {h}^{\rm rand}(x) \rangle = 0$ and
the spatial correlation between points at distance $r$ of the stochastic profile reads
\begin{equation}\label{eq:OmInfIntStochastic}
\begin{aligned}
 \left\langle h^{\rm rand}(x)\right.&\left.h^{\rm rand}(x+r) \right\rangle & \\
 =& \frac{\Delta}{\nu}\frac{L}{4\pi^2}\left[{\rm Li}_2(e^{2\pi i\frac{r}{L}})+{\rm Li}_2(e^{-2\pi i\frac{r}{L}})\right] &\\
 =& \frac{\Delta L}{2\nu} \left[\frac{r^2}{L^2}-\frac{|r|}{L}+\frac{1}{6}\right]. &
\end{aligned}
\end{equation}

\begin{figure}[h!]
\begin{center}
  \begin{tabular}{cc}
       \includegraphics[width=1\columnwidth]{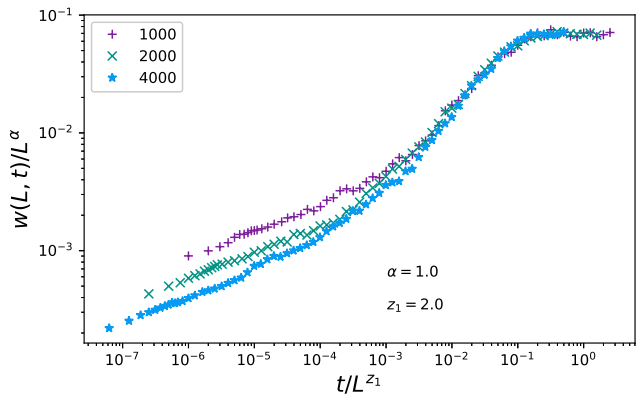}\\
  \end{tabular}
\caption{Width scaling in the $\omega\rightarrow\infty$ limit.}
\label{fig:OmInfWidthScaling}
\end{center}
\end{figure}
By squaring Eq.~(\ref{eq:OmInfInt}) and integrating over $[0,L]$, one gets the steady-state squared mean width.
This quantity comprises,  again, two competing contributions:
\begin{equation}\label{eq:OmInfSSWidth}
 w^2_{ss}(L) = c\left(\frac{\Lambda L}{2\nu}\right)^2 + \frac{\Delta L}{12\nu},
\end{equation}
the first from the squared deterministic profile, the second from the integral of $\left\langle h^{\rm rand}(x)h^{\rm rand}(x) \right\rangle$.
Here $c= \int_{-1/2}^{1/2} dx\, (x^2-|x|+1/6)^2= 1/180$.

According to Eq. (\ref{eq:OmInfSSWidth}), noise dominates the roughening dynamics for small system sizes ($L<\nu \Delta/3\Lambda^2 c$), and $\alpha=1/2$.
As the system gets larger, however, the deterministic contribution of the defect site grows in weight, until it overcomes that of the noise and sets the roughness exponent to $\alpha=1$.
As for the dynamic exponents, they can be inferred via the following argument. 
Although at early times, slope correlations spread around the defect site as in Fig. \ref{fig:CorrSpreading}, at long times one obtains diffusive behaviour.
This is because, when the tent profile has formed, growth is limited by slope diffusion: the interface must be concave ($\vee$) at the defect site for the tent to grow by one unit, so that a $+1$-slope segment has to diffuse across the $-1$-slope region on the right of the defect while a $-1$-slope segment has to diffuse across the $+1$-slope region on the left of the defect.
The resulting dynamic exponent $z_1$ equals $2$, as one would expect from the field equation (\ref{eq:EW+defect}) and is confirmed by the numerics (Fig.~\ref{fig:OmInfWidthScaling}).

The particle dynamics is also easily understood, as the particle will always be sitting on the top of the tent before moving.
Then, from Eq. (\ref{eq:PartRates}), the left and right jump rates coincide, $q_k^R = q_k^L = q$, so that the particle undergoes normal diffusion. 
As there is normal diffusion at all times, the exponents $\xi$ and $z_2$ are therefore trivially zero.
To sum up,
\begin{equation}\label{eq:OmLt1Exponents}
 \omega \gtrsim L^2: \qquad\begin{aligned}\alpha &= 1,\quad z_1=2;\\ \chi &= 0,\quad  z_2=0.\\ \end{aligned}
\end{equation}
where $\omega \gtrsim L^2$ specifically means that $\omega$ is larger than the static defect problem saturation time. 
Of course this regime can only be achieved  on a finite $L$ system and because of this the behaviour (\ref{eq:OmLt1Exponents}) does not survive the thermodynamic limit.
What happens for $\omega$ large but shorter than the saturation time is the object of the following section.

\section{The surfing regime}\label{sec:5}

\begin{figure}[h!]
\begin{center}
  \begin{tabular}{cc}
       \includegraphics[width=1\columnwidth]{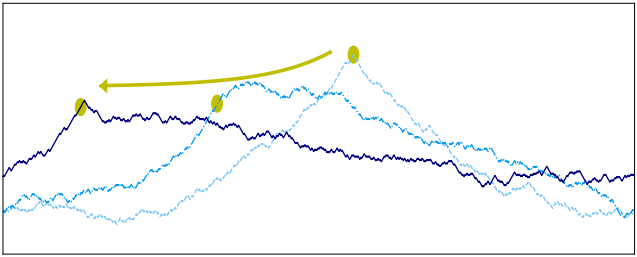}\\
  \end{tabular}
\caption{Surfing regime snapshots. The interface profiles are ordered in time according to their color, from lighter to darker, while the particle is represented by a yellow dot of a significantly larger size, to ease the understanding of the picture. The earliest snapshot (light blue), depicts the initial growth, whose dynamics is analogous to that of the $\omega\rightarrow\infty$ regime. The second (azure), is taken some moments after the particle has started moving: the wave is broken together with the left-right symmetry of the system. The last (dark blue) is the latest, and it shows that the particle keeps moving while `ironing out' the interface: this is, in fact, the regime with the smallest roughness exponent. Notice how, due to the system finite size, the particle will soon reach the back of the wave: at this point it could stochastically revert his motion, so that the long-time dynamics is still diffusive (see discussion in the text).}
\label{fig:OmLargeDynamics}
\end{center}
\end{figure}

The system behaviour when $\omega$ lies between the values of section \ref{sec:3} and \ref{sec:4} is not just a mixture of the limiting cases previously described, but another regime appears.
Such a regime---we call it the surfing regime---occurs for $\omega$ large, but still smaller than the static particle problem saturation time, which is $O(L^2)$.
The reason for the given name, as anticipated in the introduction, comes from the particle behaviour, which is peculiar to this system and specific range of parameter.
This behaviour is summarised in Fig. \ref{fig:OmLargeDynamics}.

At first, the particle pulls the interface as a static defect, thus creating the typical tent shape discussed in the previous section.
However, as $\omega$ is finite the particle will move before the tent gets as big as the system, which takes a time $O(L^2)$.
Pictorially (see Fig.~\ref{fig:OmLargeDynamics}) , the  particle randomly choses one side of the tent as the direction to move away, then the tent breaks like a sea wave towards the particle's direction of motion.
Since $\omega$ is greater than $1$, after the wave breaks the interface keeps following the particle but without completely adapting to the new particle position, as it does for $\omega\rightarrow\infty$.
As a result, the particle will keep finding itself on a downslope and being pushed forward---the particle appears to surf the interface.

The first, immediate consequence of this peculiar dynamics is that the particle is able to use the protrusive force it exerts on the membrane to propel itself.
The mechanism is qualitatively similar to the one giving rise to waves in the finite density case~\cite{cagnetta2018aa}: first the particle creates a bump, then it is advected away from it.
Although their origins lie in analogous mechanisms, the two phenomena are not quite the same, as, in the single particle system considered in this paper, directed motion is not a collective phenomenon.
Furthermore, it is not generic on the whole $\lambda,\gamma >0$ region of the parameter space, but requires $\omega$ to lie within some specific, system-size dependent values.

\begin{figure*}[ht!]
\begin{center}
  \begin{tabular}{cc}
       \includegraphics[width=1\columnwidth]{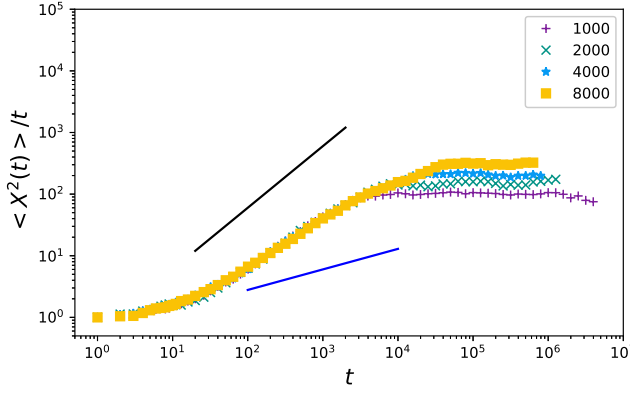} & \includegraphics[width=1\columnwidth]{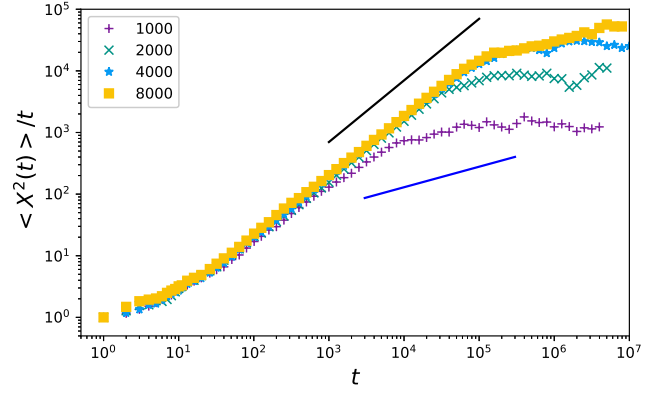}\\
  \end{tabular}
\caption{MSD at $\omega=10$ (left) and $100$ (right), system size as in the key. The dashed lines are guides to the eye, the black corresponds to  the ballistic law $\left\langle X_t^2 \right\rangle\sim t^2$ and the blue correspond to  $\left\langle X_t^2 \right\rangle\sim t^{4/3}$. Though both left and right panels display super-diffusive behaviour, 
true ballistic behaviour is achieved for $\omega=100$ only (right panel).}
\label{fig:OmLargeMSD}
\end{center}
\end{figure*}
Consider the MSDs shown in Figure \ref{fig:OmLargeMSD}.
For $\omega=10$ (left panel), after some short transient, $\left\langle X_t^2 \right\rangle$ grows faster than the $t^{4/3}$ law observed at $\omega=1$ and represented in the figure by the blue dashed line.
It is, however, slower than the black dashed lines, representing $\left\langle X_t^2 \right\rangle \sim t^2$ thus ballistic behaviour.
Upon increasing $\omega$ even further ($\omega=100$ in the right panel), the initial transient gets longer, but the superdiffusive regime becomes ballistic, at least for the bigger systems.
This regime disappears by reducing the system size simply because the crossover time to the long-time diffusive behaviour becomes short enough to mix with the initial transient.
The reason why it requires a big enough $\omega$, instead, lies in the size of the tent created before directed motion: it has to be wide enough  that it does not mix with noise-induced fluctuations and its sloped side must provide a pushing force stable against both the particle and the interface fluctuations.
For large values of $\omega$ as described, the scaling hypothesis of Eq. (\ref{eq:DiffScalingHyp}) is obeyed once again, provided one considers only systems which are big enough to display the ballistic behaviour and chooses a value of $\omega$ much larger than $1$ but still much smaller than $L^2$.

\begin{figure}[h!]
\begin{center}
  \begin{tabular}{cc}
       \includegraphics[width=1\columnwidth]{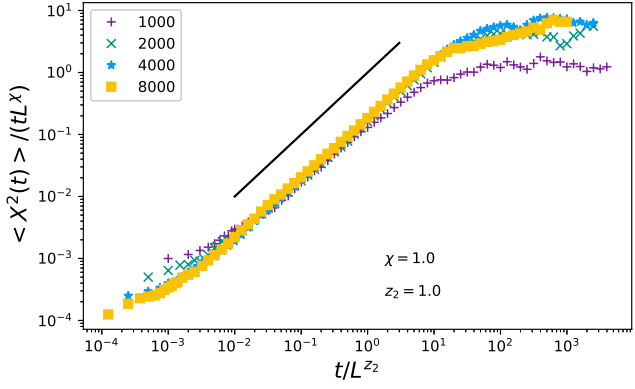}
  \end{tabular}
\caption{MSD scaling at $\omega=100$. If one excludes the $L=1000$ curve, which does not reach a full ballistic regime, the scaling exponents agree with the proposed values $\chi=z_2=1$.}
\label{fig:OmLargeMSDScaling}
\end{center}
\end{figure}
As shown in Fig. \ref{fig:OmLargeMSDScaling}, collapse is achieved for $\chi = z_2 = 1$.
The value of $z_2$, together with $\chi + z_2 = 2$, is consistent with the ballistic regime observed right before saturation.
What is the meaning of $\chi=1$? It implies that the long-time effective diffusion coefficient $D_{\rm eff}$ is directly proportional to the system size $L$.
By coupling this observation with the kinetic interpretation of the diffusion coefficient $D = (\text{mean free path})\times(\text{speed})$, we argue that the particle surfs the interface for its whole length before reverting its motion.
This is indeed what emerges by inspecting shapshots of the system as those collected in Fig. \ref{fig:OmLargeDynamics}: after travelling about a system length, the particle meets the tail of the wave it is surfing, hence it will have to stop and create a new wave to surf, possibly in the opposite direction.
The resulting motion is that of a persistent random walk, with the interface size as persistence length.

\begin{figure}[h!]
\begin{center}
  \begin{tabular}{cc}
       \includegraphics[width=1\columnwidth]{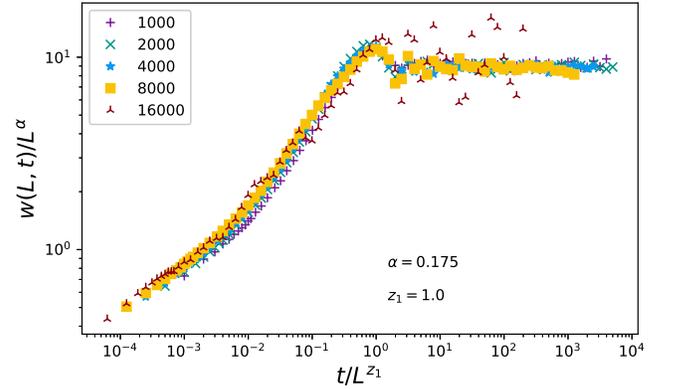}
  \end{tabular}
\caption{Width scaling at $\omega=100$. The oscillating widths collapse on a single curve for $\alpha=0.175$ and $z_1=1$.}
\label{fig:OmLargeWidthScaling}
\end{center}
\end{figure}
As a byproduct of the peculiar particle dynamics, the width scaling appears to differ from  the previous sections or indeed any of the known universality classes.
First, as in the finite density case~\cite{cagnetta2018aa}, the dynamics is dominated by oscillations: the interface roughens when the particle creates a tent, then smoothens as the particle surfs the membrane wave.
Once the particle has  stopped running,  due to the finiteness of the interface size,  the width increases again and the cycle repeats.
The period of the oscillations, being controlled by the particle running time, scales as the system size, hence $z_1=1$, as  can be observed in Fig. \ref{fig:OmLargeWidthScaling}.
Notice that this is the only case where $z_1=z_2=z_P$, as in the passive scalar problems discussed in Section~\ref{ssec:passive}.
Furthermore, due to the `ironing' performed by the surfing particle, the interface appears significantly smoother than in the other phases.
A reasonable width collapse in Fig.~\ref{fig:OmLargeWidthScaling} is achieved for $\alpha=0.175$ (about $1/6$), but we cannot exclude the possibility that the roughness will vanish upon increasing the system size even further.
Some light will be shed on this issue in the forthcoming section, where we predict the surfing regime through a self-consistent solution of a coarse-grained description of the system at hand.

\section{The Langevin description and dynamical phase transition}\label{sec:6}

In this section we resort to a coarse-grained description of our model, in order to analyse the surfing regime.
The starting point is the `Active KPZ' equation proposed in \cite{cagnetta2018aa}.
Having a single particle, though, cause the introduction of a density field to be meaningless.
We will instead build a process $X(t)$ which is a continuous-space equivalent of the jumping particle, at least to the extent at which the height field $h(x,t)$ is the continuous-space equivalent of our discrete interface.

Hence, let us start by calling $i_t$ the lattice position of the particle at time $t$, $a$ the lattice spacing, and define $x = a i$ as the coordinate that will become continuous in the $a\rightarrow 0$ limit.
The current particle position will be distinguished from the latter by denoting it with $X_t$.
After a short time $\delta t$, the particle postion changes according to
\begin{equation}
  X_{t+\delta t} = \left\lbrace\begin{aligned} &X_t + a,&\text{ prob. }\delta t q_i^R,\\&X_t - a,&\text{ prob. }\delta t q_i^L, \\&X_t,&\text{ otherwise.} \\\end{aligned}\right.
\end{equation}
Thus,
\begin{equation}
\begin{aligned}
 \left\langle \delta X_t \right\rangle &= \delta t a \left(q_i^R-q_i^L\right) = -\delta t a 2q \gamma \nabla h_i;\\
 \left\langle \delta X_t^2 \right\rangle &= \delta t a^2 \left(q_i^R+q_i^L\right) = \delta t a^2 2q ,\\
\end{aligned}
\end{equation}
and higher order moments $\left\langle \delta X_t^n \right\rangle$ are of order $a^n$.
By setting $2q=a^{-2}$, approximating $\nabla h_i$ with $a\partial_x h(x,t) + O(a^2)$ and performing the $a\rightarrow 0$ limit, one notices that the $n$-th moments with $n>2$ vanish, while
\begin{equation}
 \frac{\left\langle \delta X_t \right\rangle}{\delta t} = -\gamma \partial_x h(x,t)|_{X_t};\quad \frac{\left\langle \delta X_t^2 \right\rangle}{\delta t} = 1.
\end{equation}
The quantities above are nothing but the first two coefficients of the Kramers-Moyal expansion of the probability distribution of the particle position.
From those we deduce the contiuous-space limit particle postion obeys the Langevin equation
\begin{equation}\label{eq:LanFieldModela}
\dot{X}_t = -\gamma \partial_x h(x,t)|_{X_t} + \xi(t),
\end{equation}
where $\xi(t)$ is a Gaussian white noise with unit variance.
With an analogous procedure one can derive the height equation (cf. Eq. (\ref{eq:EW+defect}))
\begin{equation}\label{eq:LanFieldModelb}
 \partial_t h = \omega\left[\Lambda\delta(x-X_t) + \nu\nabla^2 h\right] + \sqrt{2\Delta\omega}\eta,
\end{equation}
where we factored out the timescale ratio parameter $\omega$.

Without  solving Eq.~(\ref{eq:LanFieldModela},\ref{eq:LanFieldModelb}) explicitly, we can check if it admits a surfing solution in the deterministic limit $\Delta=0$.
Let us work, as in section \ref{sec:4}, in the Fourier representation: with an initially flat interface and $\Delta=0$,
\begin{equation}\label{eq:IntModes}
 \tilde{h}_k(t) =  \omega\Lambda\int_0^t ds\,e^{-\omega\nu k^2(t-s)} e^{-ikX_s}.
\end{equation}
As he speed of the slider equals the force felt at time $t$, $F(t) = -\gamma\partial_x h(x,t)|_{X_t}$,
\begin{equation}\label{eq:MemLan}
\begin{aligned}
 \dot{X}_t &= -\gamma\int_{-\infty}^{\infty}\frac{dk}{2\pi}\, ik \tilde{h}_k(t) e^{ikX_t}  \\
 & = - \omega\Lambda\gamma\int_{-\infty}^{\infty}\frac{dk}{2\pi}\,\int_0^t ds\,ik e^{-\omega\nu k^2(t-s)} e^{ik(X_t-X_s)}.
\end{aligned}
\end{equation}
The above equation is just a reformulation of Eq.~(\ref{eq:LanFieldModela},\ref{eq:LanFieldModelb}), obtained by integrating out the height field.
Since the particle begins its motion on a flat interface, $\dot{X}_t$ vanishes at $t=0$.
$\dot{X}_t =0$ is actually a  solution of Eq.~(\ref{eq:MemLan}) at \textit{all} times, as can be checked by setting $X_t=X_s$ in the equation right-hand side---the ensuing parity of the integrand cause the $k$-integral to vanish.

Our simulations, however, show that the particle starts moving at some later time.
Furthermore, we have argued that the run length  diverges in the thermodynamic limit $L\rightarrow\infty$ (cf. section \ref{sec:5}), which should imply the existence---and finiteness---of the following limit,
\begin{equation}\label{eq:SCA1}
 \lim_{t\rightarrow\infty} \dot{X}_t = v_{ss}
\end{equation}
where we refer to $v_{ss}$ as a 'steady-state' surfing speed.
The goal of the remainder of the section is then that of finding a self-consistent equation for $v_{ss}$ by performing a long-time limit of Eq.~(\ref{eq:MemLan}).
This  limit is given by,
\begin{equation}\label{eq:SCA2}
\begin{aligned}
v_{ss} = \lim_{t\rightarrow\infty} \left\lbrace -\omega\Lambda\gamma\int_{-\infty}^{\infty}\frac{dk}{2\pi}\,\int_0^t ds\,ik e^{\left[ikv_{ss}-\omega\nu k^2\right](t-s)}\right\rbrace,
\end{aligned}
\end{equation}
In Eq.~(\ref{eq:SCA2}), $X_t-X_s$ is approximated by $v_{ss}(t-s)$ for all times whereas actually $v_{ss}$ approximates $\dot{X}_t$ only at very large times.
However, due to the decaying exponential factor $\exp{\{-\omega\nu k^2(t-s)\}}$, the $s$-integral is insensitive to the integrand values for small $s$ when $t\rightarrow\infty$.
Therefore, under the $t\rightarrow\infty$ limit, the replacement of $X_t-X_s$ with $v_{ss}(t-s)$ can be safely extended to the whole integration domain $[0,t]$ to get to Eq.~(\ref{eq:SCA2}).
We finally get to the sought self-consistent equation by performing the  limit.
The result reads
\begin{equation}\label{eq:SCEq}
\begin{aligned}
 v_{ss} &= \frac{\gamma\Lambda}{\nu}\int_{-\infty}^{\infty}\frac{dk}{2\pi}\, \frac{1}{ik + v_{ss}/\omega\nu}  \\
 &= \frac{\gamma\Lambda}{2\nu} \sign(v_{ss}),
\end{aligned}
\end{equation}
where $\sign$ is the signum function ($+1$ for positive argument, $-1$ for negative, $0$ when the argument vanishes).
Eq.~(\ref{eq:SCEq}) is solved by $v=0$ and $\pm v^*$, where $v^* = \gamma\Lambda/2\nu$.

This means that, at some time between $0$ and $\infty$, the two non-zero solutions appear: in order to infer this time, one would have to solve the full time-dependent problem.
The time-dependent particle speed will then interpolate from $v=0$ at $t=0$ to $v = \gamma\Lambda/2\nu \sign(v)$ at $t\rightarrow\infty$.
The approach to the $t\rightarrow\infty$ limit follows easily from next-to-leading corrections to Eq.~(\ref{eq:SCA2}), found to be exponentially small in $t$ and controlled by the saturation time $t_{sat}=4\omega\nu/v^*$.
The departure from the $t=0$ solution, instead, is much harder to analyse.
The reason is that a small $t$ expansion of Eq.~(\ref{eq:MemLan}) will result in all the derivatives of $X_t$ vanishing. 
We believe this might be due to some \textit{latency period} where the solution sticks to $\dot{X}_t =0$.

\begin{figure}[h!]
\begin{center}
  \begin{tabular}{cc}
       \includegraphics[width=1\columnwidth]{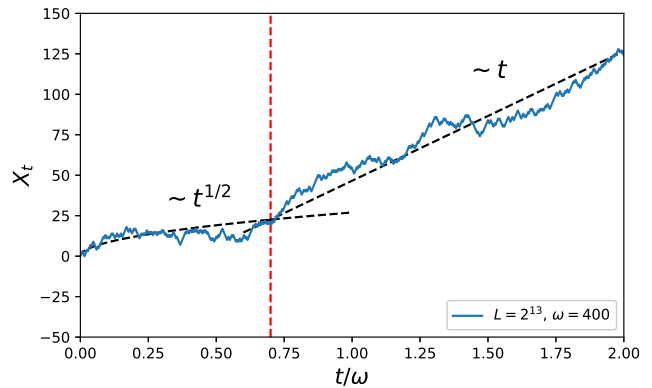}
  \end{tabular}
\caption{Particle trajectory for $\omega=400$, with $L$ as in the key. The depicted behaviour is emblematic of the whole $1\ll\omega\lesssim L^2$ regime. The dynamics goes qualitatively as follows. Starting from a flat interface at $t=0$, the particle fluctuates diffusively. It eventually picks a direction at random and starts running, but it does so only after some finite time.}
\label{fig:LatencyTime}
\end{center}
\end{figure}
As the transition away from this solution has proven  difficult to analyse, we present in Fig.~\ref{fig:LatencyTime} a stochastic simulation that illustrates it.
The transition to a running phase with constant speed occurs after some apparent latency period, marked by a vertical red dashed line in the figure. 
In this simulation the particle is subject to thermal noise when there is no net slope driving it, hence it moves diffusively before making the transition to the running phase.
It is possible that some thermal kick is also required for the particle to transition between the $v=0$ and $v=\pm v^*$ deterministic solutions.

\section{Conclusions}
\label{sec:conc}

In conclusion, we have studied the statistical mechanics of a single active particle -- an active slider -- on a fluctuating membrane.
The nature of the coupling is such that the active particle stimulates interfacial growth, and is in turn affected by height gradients so as to be repelled by peaks and slide down to accumulate at valleys.
We chose this setup for two reasons.
First,  the particles create interface peaks which repel them, thus perpetually generating activity and this enhances the non-equilibrium nature of the problem (i.e., there is no equilibrium system which qualitatively resembles the one we study).
Second, this setup may be relevant to understanding the behaviour of membrane proteins, especially those which, by signalling to the actin cortex, stimulate membrane growth.

Our simulations show that there is a surprisingly rich range of possible dynamical behaviours of an active slider.
The regimes we identified depend crucially on the ratio between two timescales, that of interface relaxation and that of slider motion/diffusion in an effective potential.
When the first timescale is sufficiently slow, the interface behaves as an Edwards-Wilkinson equilibrium interface, whereas the particle dynamics is non-trivial: it is superdiffusive at intermediate times, and diffusive at late times.
For very slow interfacial motion, the slider can also move subdiffusively at very early times, which is reminiscent of the dynamics of random walkers in quenched random environments.
Thus the interface and the slider exhibit different dynamic exponents.
It is intriguing to notice that, in the slow interface regime, the slider dynamics resembles metadynamics, a virtual dynamics used to let a system fully sample a possibly complex free energy landscape---it would be tempting here to speculate that such metadynamics could also appear as an efficient strategy in biological systems.

When, instead, the dynamics of the interface is fast with respect to particle diffusion, two additional distinct regimes can occur, whose interplay depends also on the system size.
One is the ``surfing'' regime, where the slider travels ballistically for whole system lengths by riding its own wave, the other is the ``electrostatic'' regime where the slider behaves as a moving positive charge on a negatively charged ring.
They both occur due to the particle being able, if slow enough, to enslave the interface: after an adequate amount of time, the interface modification due to the slider action will dominate over thermal fluctuations, and the interface will look like the tent described in Section \ref{sec:4}.
Depending on whether the slider influences the whole interface or part of it before it moves, the tent-like shape will stay stable or break and be surfed by the slider.
Such an added system-size dependence leads us to believe that, should the $L\rightarrow\infty$ limit be performed first, only the metadynamics and surfing regimes would survive. Which of the two takes place would then be determined by the relative importance of noise- and activity-induced shape fluctuations.

A further exploration of the effect of noise would be of great interest.
One might, for instance, include noise in the coarse-grained approach we presented in Section \ref{sec:6} (coupled Langevin Eq.~(\ref{eq:LanFieldModela},\ref{eq:LanFieldModelb})).
Arguably, the addition of interfacial noise could hinder the emergence of surfing solutions, hence explain why surfing only appears in a specific range of $\omega$.
Furthermore, as the interface noise influences all the particles in the system, its inclusion would consitute a significant step in understanding the interface-mediated interactions between the sliders and the emergent collective behaviour.
Even the simpler---but still challenging---exact solution of the deterministic problem, Eq.~(\ref{eq:MemLan}), would shed some light on the transition to directed motion described in Section \ref{sec:6}.

To sum up, our work shows that non-equilibrium active membranes exhibit non-trivial dynamics, even when activity is due to the action of a single particle. We hope that this work will stimulate further theoretical studies of nonequilibrium membranes with active sliders, as well as other applications of this models to other contexts, for instance that of nonequilibrium random walkers in fluctuating potentials, or chemotactic microorganisms and phoretic particles~\cite{liebchen2017phoretic}. 

FC acknowledges support from SFC under a studentship.

\bibliographystyle{apsrev4-1.bst}
\bibliography{PartFlucField,KPZ,PRE2018}

\end{document}